\documentclass{elsart}
\journal{ \bf This is a draft version dated\ }

\renewcommand{\theequation}{\thesection.\arabic{equation}}
\usepackage[centertags]{amsmath}
\usepackage{newlfont}

\usepackage{graphicx}
\usepackage{psfrag}

\begin{document}
\begin{frontmatter}
\title{The homotopy analysis method and the Li\'{e}nard equation}

\author[Tehran,Ghazvin]{S. Abbasbandy\corauthref{cor}},
\ead{abbasbandy@yahoo.com}
\author[Pamp]{J.L. L\'{o}pez},
\author[Zara]{R. L\'{o}pez-Ruiz}

\corauth[cor]{Corresponding author.}

\address[Tehran]{Department of Mathematics, Science and Research
 Branch, Islamic Azad University, Tehran, 14778, Iran}

\address[Ghazvin]{Department of Mathematics, Imam Khomeini
International University, Ghazvin, 34149-16818, Iran}

\address[Pamp]{Department of Mathematical and Informatics Engineering,
  Universidad P\'ublica de Navarra, 31006-Pamplona, Spain}

\address[Zara]{Department of Computer Science and BIFI,
  Universidad de Zaragoza, 50009-Zaragoza, Spain}

\date{\today}

\begin{abstract}
In this work, Li\'enard equations are considered.
The limit cycles of these systems are studied by applying
the homotopy analysis method. The amplitude and frequency obtained with this methodology
are in good agreement with those calculated by computational methods.
This puts in evidence that the homotopy analysis method is an useful tool
to solve nonlinear differential equations.
\end{abstract}

\end{frontmatter}

\section{Introduction}
\setcounter{equation}{0}

A generalization of the van der Pol oscillator is the classical
Li\'{e}nard differential equation,
\begin{equation}
\ddot{x}(t) + \epsilon f(x) \dot{x}(t) + x(t) =0,\ \ \ t \geq 0,
\label{lie}
\end{equation}
with $\epsilon$ a real parameter and $f(x)$ any real function.
The dot denotes the derivative with respect to time $t$.
The periodic solutions of this system are called {\it limit cycles} \cite{andronov}.
For instance, when $f(x)=x^2-1$ (van der Pol oscillator),
Eq. \ref{lie} displays a limit cycle whose uniqueness and non-algebraicity
has been shown for the whole range of the parameter $\epsilon$ \cite{odani}.
Its behavior runs from near-harmonic oscillations when $\epsilon\rightarrow 0$
to relaxation oscillations when $\epsilon\rightarrow\infty$, making
it a good model for many practical situations \cite{lopezruiz}.
Other partial results on the number and form of limit cycles in
Li\'enard systems are scattered in the literature \cite{ye}.
When $f(x)$ is a polynomial of degree $N=2n+1$ or $2n$, with $n$ a
natural number, Lins, Melo and Pugh have conjectured
(LMP-conjecture) that the maximum number of limit cycles allowed
is just $n$ \cite{lmp}. It is true if $N = 2$, or $N = 3$ or if
$f(x)$ is even and $N = 4$ \cite{lmp,ryc}. Also, there are strong
arguments for claiming its truth in the strongly nonlinear regime
$(\epsilon\rightarrow \infty)$ when $f(x)$ is an even polynomial
\cite{ll1} and recently in the weakly nonlinear regime
$(\epsilon\rightarrow 0)$ for even $f(x)$ \cite{ll2}.
However, this conjecture has been recently
shown \cite{dumortier} to have counterexamples for $n\geq 3$ when $f(x)$ is not even.
In particular, it has been found a polynomial $f(x)$ of degree $6$
such that the associated Li\'enard equation has at least $4$ limit cycles \cite{dumortier}.
Thus, there are no general results
about the limit cycles when $f(x)$ is a polynomial
of degree greater than $5$ neither, in general, when $f(x)$
is an arbitrary real function \cite{gn}.

Apart from the classical perturbative techniques that can be applied in the weakly
nonlinear regime \cite{verhulst,anderson,ll3},
different non-perturbative approaches allowing to obtain information on
the number of limit cycles and their location in phase space have
been proposed in the last years.
A method that gives a sequence of algebraic approximations to the equation
of each limit cycle can be found in \cite{gn}, and a variational method
showing that limit cycles correspond to relative extrema of certain functionals
is explained in \cite{depassier}. Here, we are interested in the
application of another non-perturbative technique,
{\it the homotopy analysis method} (HAM), to this problem.
Liao \cite{li1,li2} has developed this purely analytic technique
to solve nonlinear problems in science and engineering.
The HAM has been applied successfully to
many nonlinear problems such as free oscillations of self-excited systems
\cite{li3}, the generalized Hirota--Satsuma coupled KdV
equation \cite{ab2}, heat radiation \cite{ab3}, finding the
root of nonlinear equations \cite{ab6}, finding solitary-wave
solutions for the fifth-order KdV equation \cite{ab4}, finding
solitary wave solutions for the Kuramoto--Sivashinsky equation
\cite{ab5}, finding the solitary solutions for the Fitzhugh-Nagumo
equation \cite{ab7}, boundary-layer flows over an impermeable
stretched plate \cite{li4}, unsteady boundary-layer flows over a
stretching flat plate \cite{li7}, exponentially decaying boundary
layers \cite{li6}, a nonlinear model of combined convective and
radiative cooling of a spherical body \cite{li5}, and many other
problems (see \cite{txl,WL,ha4,ha5,lc,sa,ta,wa,AP1}, for example).

In this paper, we are interested in applying the HAM to Li\'enard equation
(\ref{lie}) in order to obtain good approximations to the amplitude and shape of its
limit cycles. These calculations are explained in Section 2.
The validity of the method (for arbitrary $\epsilon$)
is shown for the different particular cases analyzed in Section 3.
Last section includes our conclusions.

\section{HAM applied to Li\'enard equations}
\setcounter{equation}{0}

In general, the limit cycles of (\ref{lie}) contain
two important physical parameters, i.e. the frequency $\omega$ and
the amplitude $a$. So, without loss of any generality, consider
such initial conditions:
\begin{equation} \label{cond}
x(0)=a,\ \ \ \dot{x}(0)=0,
\end{equation}
where $a>0$ is the amplitude of the limit-cycle.

Let $\tau=\omega t$ denotes a new time scale, with $\omega>0$.
Under the transformation
\begin{equation} \label{trans}
\tau=\omega t,\ \ \ x(t)=au(\tau),
\end{equation}
the original Eq. (\ref{lie}) and its initial conditions
(\ref{cond}) become
\begin{equation} \label{lien}
\omega^2 u''(\tau) + \epsilon \omega f(au) u'(\tau) + u(\tau)=0,
\end{equation}
and
\begin{equation} \label{condn}
u(0)=1,\ \ \ u'(0)=0,
\end{equation}
respectively, where the prime denotes the derivative with respect
to $\tau$.

The limit-cycles of (\ref{lien}) are periodic motions with period
$T=2\pi/\omega$ and thus $u(\tau)$ can be expressed by
\begin{equation} \label{expe}
u(\tau)=\sum_{m=0}^{+\infty} \big[\alpha_m \sin(m\tau)+\beta_m
\cos(m\tau)\big],
\end{equation}
where $\alpha_m$ and $\beta_m$ are coefficients to be determined.
According to the {\it rule of solution expression} denoted by
(\ref{expe}) and the boundary conditions (\ref{condn}), it is
natural to choose
\begin{equation}
u_0(\tau)=\cos(\tau), \label{u0}
\end{equation}
as the initial approximation to $u(\tau)$. Let $\omega_0$ and
$a_0$ denote the initial approximations of the frequency $\omega$
and the amplitude $a$, respectively.

We define an auxiliary linear operator ${\cal L}$ by
\begin{equation} \label{eq:24}
{\cal L} [\phi(\tau;p)] =  \omega_0^2 \left(\frac
{\partial^2}{\partial \tau^2} + 1 \right) \phi(\tau;p),
\end{equation}
with the property
\begin{equation} \label{eq:25}
{\cal L} [C_1 \sin(\tau) + C_2 \cos(\tau)]=0,
\end{equation}
where $C_1$ and $C_2$ are constants, and $p$ is a parameter explained below.

From (\ref{lien}) we define a nonlinear operator
\begin{equation} \label{nlo}
 {\cal N}[\phi(\tau;p),A(p),\Omega(p)]=
\Omega^2(p) \frac{\partial^2 \phi(\tau;p)}{\partial \tau ^2} +
\epsilon \Omega(p) f (A(p) \phi(\tau;p)) \frac{\partial
\phi(\tau;p)}{\partial \tau} + \phi(\tau;p),
\end{equation}
and then construct the homotopy
\begin{equation} \label{hom}
{\cal H}[\phi(\tau;p),A(p),\Omega(p)]=(1-p){\cal L}
[\phi(\tau;p)-u_0(\tau)] -h p {\cal
N}[\phi(\tau;p),A(p),\Omega(p)],
\end{equation}
where $h$ is a nonzero auxiliary parameter. Setting ${\cal
H}[\phi(\tau;p),A(p),\Omega(p)]=0$, we have the zero-order
deformation equation
\begin{equation} \label{zod}
(1-p){\cal L} [\phi(\tau;p)-u_0(\tau)]=h p {\cal N}
[\phi(\tau;p),A(p),\Omega(p)],
\end{equation}
subject to the boundary conditions
\begin{equation} \label{phiz}
\phi(0;p)=1,\qquad\frac{\partial\phi(\tau;p)}{\partial\tau}\Big|_{\tau=0}=0,
\end{equation}
where $p\in [0,1]$ is an embedding parameter. When the parameter
$p$ increases from 0 to 1, the solution $\phi(\tau;p)$ varies from
$u_0(\tau)$ to $u(\tau)$, $A(p)$ varies from $a_0$ to $a$, and
$\Omega(p)$ varies from $\omega_0$ to $\omega$. Assume that
$\phi(\tau;p),\ A(p)$ and $\Omega(p)$ are analytic in $p\in[0,1]$
and can be expanded in the Maclaurin series of $p$ as follows:
\begin{equation} \label{mac}
\phi(\tau;p)=\sum_{m=0}^{+\infty} u_m(\tau)p^m,\ \
A(p)=\sum_{m=0}^{+\infty} a_m p^m,\ \
\Omega(p)=\sum_{m=0}^{+\infty} \omega_m p^m,\
\end{equation}
where
\begin{equation*}
u_m(\tau)=\frac{1}{m!} \frac{\partial ^m \phi(\tau;p)}{\partial
p^m}\Big|_{p=0},\qquad a_m=\frac{1}{m!} \frac{\partial^m
A(p)}{\partial p^m}\Big|_{p=0},\qquad \omega_m=\frac{1}{m!}
\frac{\partial^m \Omega(p)}{\partial p^m}\Big|_{p=0}.
\end{equation*}
Notice that series (\ref{mac}) contain the auxiliary parameter
$h$, which has influence on their convergence regions. Assume that
$h$ is properly chosen such that all of these Maclaurin series
are convergent at $p=1$. Hence at $p=1$ we have
\begin{equation*}
u(\tau)=u_0(\tau)+\sum_{m=1}^{+\infty} u_m(\tau),\qquad
a=a_0+\sum_{m=1}^{+\infty} a_m,\qquad
\omega=\omega_0+\sum_{m=1}^{+\infty} \omega_m.
\end{equation*}
At the $M$th-order approximation, we have the analytic solution of
Eq.~(\ref{lien}), namely
\begin{equation} \label{eq:14}
u(\tau) \approx U_M(\tau)=\sum_{m=0}^M u_m(\tau),\ \ a \approx A_M
= \sum_{m=0}^M a_m,\ \ \omega \approx \Omega_M = \sum_{m=0}^M
\omega_m.
\end{equation}
The auxiliary parameter $h$ can be employed to adjust the
convergence region of the series (\ref{eq:14}) in the homotopy
analysis solution. By means of the so-called $h$-curve, it is
straightforward to choose an appropriate range for $h$ which
ensures the convergence of the solution series. As pointed out by
Liao \cite{li2}, the appropriate region for $h$ is indicated when
$a$ and $\omega$ are horizontal segments when plotted versus $h$.

Differentiating Eqs.~(\ref{zod}) and (\ref{phiz}) $m$ times with
respect to $p$, then setting $p=0$, and finally dividing by
$m!\,$, we obtain the $m$th-order deformation equation
\begin{equation} \label{mod}
{\cal L} [u_m(\tau)-\chi_m u_{m-1}(\tau)]=h R_m(\tau), \qquad
(m=1,2,3,\ldots),
\end{equation}
subject to the boundary conditions
\begin{equation} \label{zmbcs}
u_m(0)=0,\quad u_m'(0)=0,
\end{equation}
where $R_m(\tau)$ is defined by
\begin{equation} \label{eq:11}
R_m(\tau)= {1 \over (m-1)!} {\partial ^{m-1} {\cal
N}[\phi(x;p),A(p),\Omega(p)] \over
\partial p^{m-1}} \Big|_{p=0},
\end{equation}
and
\begin{equation*}
\chi_m= \left\{ \begin{array}{ll} 0, & m\leq 1, \\ 1, & m>1.
\end{array} \right.
\end{equation*}

Notice that, both $a_m$ and $\omega_m$ remain unknown and due to
the form of the solution (\ref{expe}) and definition
(\ref{eq:24}), solutions of (\ref{mod}) and (\ref{zmbcs}) should
not contain the secular terms $\tau\sin(\tau)$ and
$\tau\cos(\tau)$. It is easy to check that ${\cal L} [t\sin t]=2\cos t$
and ${\cal L} [t\cos t]=-2\sin t$, then the right-hand side term $R_m(\tau)$ of
(\ref{eq:11}) should not contain the terms $\sin(\tau)$ and
$\cos(\tau)$ in order to avoid the secular terms in the solution.
Hence, the coefficients of $\sin(\tau)$ and
$\cos(\tau)$ must be zero. If we rewrite
\begin{equation*}
R_m(\tau)= \sum_{i=1}^{\psi(m)} \big[ c_{m,i} \cos(i\tau) +
d_{m,i} \sin(i\tau) \big],
\end{equation*}
then
\begin{equation*}
c_{m,i}={2 \over \pi} \int_0^\pi R_m(\tau) \cos(i\tau) \text{d}
\tau, \ \ \ d_{m,i}={2 \over \pi} \int_0^\pi R_m(\tau) \sin(i\tau)
\text{d} \tau,
\end{equation*}
become zero when $i>\psi(m)$. Hence, we have two algebraic
equations
\begin{equation} \label{ale}
c_{m,1}=0, \qquad d_{m,1}=0,
\end{equation}
which determine $a_{m-1}$ and $\omega_{m-1}$ for $m=1,2,3,\ldots$.
The above two algebraic equations are often non-linear for $a_0$
and $\omega_0$ when $m=1$, but always linear in other case, as
proved by Liao \cite{li1}. So, after solving $a_{m-1}$ and
$\omega_{m-1}$, it is easy to gain the solution of (\ref{mod}) and
(\ref{zmbcs}) as
\begin{equation} \label{sol}
u_m(\tau) = \chi_m u_{m-1}(\tau) + \sum_{i=2}^{\psi(m)}
 \frac{c_{m,i}\cos(i\tau)+d_{m,i} \sin(i\tau)}{\omega_0^2 (1-i^2)}
  + C_1 \cos(\tau) + C_2 \sin(\tau),
\end{equation}
where the coefficients $C_1$ and $C_2$ are determined by
(\ref{zmbcs}). In this way, one can gain $a_{m-1}$, $\omega_{m-1}$
and $u_m(\tau)$ for $m=1,2,3,\ldots$, successively.

\section{Some examples}
\setcounter{equation}{0}

In this section, the validity of the proposed method is
illustrated by two examples. The limit cycles of different families of Li\'{e}nard systems
were studied in the weakly nonlinear regime \cite{ll2,ll4}.

\noindent\textbf{Example 1.} The van der Pol oscillator is defined
for $f(x)=x^2-1$. This system has a unique limit cycle, which is
stable for $\epsilon>0$.

The corresponding perturbation approximation of the amplitude
gives by a recursive algorithm the following formula
\begin{equation} \label{lla}
a(\epsilon)=2+{1\over 96}\epsilon^2-{1033\over 552960}\epsilon^4+
{1019689\over 55738368000}\epsilon^6+\cal{O}(\epsilon^8),
\end{equation}
reported in \cite{ll2,ll3}. This analytical result agrees for small $\epsilon$ with
the computational calculation of the `exact' amplitudes calculated by a fourth-order
Runge-Kutta method. Also, the expansion in $\epsilon$ of the frequency was obtained
in \cite{anderson} up to order $\cal{O}(\epsilon^{24})$. For simplicity we give the
expansion up to order $\cal{O}(\epsilon^{8})$:
\begin{equation} \label{ander-geer}
\omega(\epsilon) =  1 - \frac{{\epsilon }^2}{16} +
\frac{17\,{\epsilon }^4}{3072} + \frac{35\,{\epsilon }^6}{884736}
+\cal{O}(\epsilon^8).
\end{equation}

Under transformation (\ref{trans}), Eq. (\ref{lie}) becomes
\begin{equation} \label{lie1}
\omega^2 u''(\tau) + \epsilon \omega \big[ a^2 u^2(\tau)-1 \big]
u'(\tau) + u(\tau)=0.
\end{equation}

From (\ref{eq:11}), the term $R_m(\tau)$ in (\ref{mod}) becomes
\begin{eqnarray} \label{rm1}
R_m(\tau) & = & \sum_{n=0}^{m-1} u''_{m-1-n}(\tau) \Big(
\sum_{j=0}^n \omega_j \omega_{n-j} \Big) + u_{m-1}(\tau) -
\epsilon \sum_{n=0}^{m-1} \omega_n u'_{m-n-1}(\tau) \\
& & + \epsilon \sum_{n=0}^{m-1}\Big[ \Big( \sum_{i=0}^{m-1-n}
\omega_i u'_{m-n-i-1}(\tau) \Big) \sum_{j=0}^n \Big( \sum_{r=0}^j
a_r a_{j-r} \Big) \Big( \sum_{s=0}^{n-j} u_s(\tau) u_{n-j-s}(\tau)
\Big)\Big].  \nonumber
\end{eqnarray}

It is found that the frequency $\omega$ and the amplitude $a$ at
the $M$th-order of approximation can be expressed by
\begin{equation} \label{res1}
\omega \approx \Omega_M = \omega_0 + \sum_{i=1}^M \epsilon^{2i}
\sum_{j=i}^M
 \alpha_M^{i,j} h^j,\ \ \
a \approx A_M = a_0 + \sum_{i=1}^{M-1} \epsilon^{2i}
\sum_{j=i+1}^M
 \beta_M^{i,j} h^j,
\end{equation}
respectively. So, $a_0$ and $\omega_0$ are obtained by solving
(\ref{ale}) for $m=1$, i.e.
\begin{equation*}
c_{1,1}=(1-\omega_0^2)=0,\ \ \ d_{1,1}=\epsilon \omega_0
(1-{1\over 4} a_0^2)=0.
\end{equation*}
Hence, we have unique limit cycle by $\omega_0=1$ and $a_0=2$.

Note that results (\ref{res1}) contain the auxiliary parameter
$h$. It is found that convergence regions of the approximation
series are dependent upon $h$. The obtained results for
amplitude are as follows
\begin{eqnarray*}
A_1 & = & 2, \\
A_2 & = & 2 + {h^2 \over 96} \epsilon^2, \\
A_3 & = & 2 + \frac{h^2\,{\epsilon }^2}{32} + \frac{h^3\,{\epsilon
}^2}{48} + \frac{h^3\,{\epsilon }^4}{768}, \\
A_4 & = & 2 + \frac{h^2\,{\epsilon }^2}{16} + \frac{h^3\,{\epsilon
}^2}{12} + \frac{h^4\,{\epsilon }^2}{32} +
  \frac{h^3\,{\epsilon }^4}{192} + \frac{1847\,h^4\,{\epsilon }^4}{552960} + \frac{h^4\,{\epsilon
  }^6}{6144},
\end{eqnarray*}
and for frequency are
\begin{eqnarray*}
\Omega_1 & = & 1 + {h \over 16} \epsilon^2, \\
\Omega_2 & = & 1 + \frac{h\,{\epsilon }^2}{8} +
\frac{h^2\,{\epsilon }^2}{16} + \frac{3\,h^2\,{\epsilon }^4}{512},
\\
\Omega_3 & = & 1 + \frac{3\,h\,{\epsilon }^2}{16} +
\frac{3\,h^2\,{\epsilon }^2}{16} + \frac{h^3\,{\epsilon }^2}{16} +
  \frac{9\,h^2\,{\epsilon }^4}{512} + \frac{37\,h^3\,{\epsilon }^4}{3072} + \frac{5\,h^3\,{\epsilon
  }^6}{8192}, \\
\Omega_4 & = & 1 + \frac{h\,{\epsilon }^2}{4} +
\frac{3\,h^2\,{\epsilon }^2}{8} + \frac{h^3\,{\epsilon }^2}{4} +
  \frac{h^4\,{\epsilon }^2}{16} + \frac{9\,h^2\,{\epsilon }^4}{256} + \frac{37\,h^3\,{\epsilon }^4}{768} +
  \frac{19\,h^4\,{\epsilon }^4}{1024} \\ & & + \frac{5\,h^3\,{\epsilon }^6}{2048} + \frac{95\,h^4\,{\epsilon }^6}{49152} +
  \frac{35\,h^4\,{\epsilon }^8}{524288}.
\end{eqnarray*}
For example, for $h=-1$, the 10th-order approximation gives
\begin{eqnarray*}
A_{10} & = & 2 + \frac{{\epsilon }^2}{96} - \frac{1033\,{\epsilon
}^4}{552960} + \frac{1019689\,{\epsilon }^6}{55738368000} +
  \frac{9835512276689\,{\epsilon }^8}{157315969843200000} - \\ & &
  \frac{58533181813182818069\,{\epsilon }^{10}}{7326141789209886720000000}
  + \cal{O}(\epsilon^{12}), \\
\Omega_{10} & = & 1 - \frac{{\epsilon }^2}{16} +
\frac{17\,{\epsilon }^4}{3072} + \frac{35\,{\epsilon }^6}{884736}
-  \frac{678899\,{\epsilon }^8}{5096079360} +
\frac{28160413\,{\epsilon }^{10}}{2293235712000} +
  \cal{O}(\epsilon^{12}).
\end{eqnarray*}

The general solution of Eq.~(\ref{mod}) is
\begin{equation}
u_m(\tau)= \hat{u}_m(\tau) + C_1 \sin(\tau) + C_2 \cos(\tau),
\label{zmgen}
\end{equation}
where $C_1$ and $C_2$ are constants and $\hat{u}_m(\tau)$ is a
particular solution of Eq.~(\ref{mod}). Using ({\ref{zmbcs}), we
can obtain the unknowns $C_1$ and $C_2$.

Our solution series contain the auxiliary parameter $h$. We
can choose appropriate value of $h$ to ensure that the three
solution series (\ref{eq:14}) converge. We can investigate the
influence of $h$ on the convergence of $a$ and $\omega$ by
plotting the curve of $a$ and $\omega$ versus $h$, as shown in
Figs.~1 and 2. One can see on these plots that, for $\epsilon=1$,
we have $-1.4\leq h \leq -0.4$ and for $\epsilon=0.5$, we have
$-1.4\leq h \leq -0.2$. The comparison of the amplitude $a$
and the frequency $\omega$ at the 10th-order of approximation with
the numerical results is as shown in Figs.~3 and 4, where
$h=-1,\ -{2\over 3}$ and $-{1\over 3}$. However, as $h$ is
negative and close to zero, the convergence region becomes larger
and larger. Note that, one has a great freedom to choose the
auxiliary parameter $h$. Certainly, this can be chosen as a
function of $\epsilon$. Due to (\ref{res1}), the frequency and the
amplitude are even functions of $\epsilon$. Hence, $h$ should
be an even function of $\epsilon$. For example, we can take $h=-{1\over
\sqrt{1+\gamma \epsilon^2}}$, where $\gamma$ is a positive
constant. As $\gamma$ increases, the convergence regions of the
amplitude and the frequency become larger and larger, as shown in
Figs.~5 and 6.

We can integrate Eq.~(\ref{lie}) by Runge-Kutta method in order to
obtain the limit cycle and its properties. Table 1 shows the
value of the amplitude $a_{RK}$ obtained by using Runge-Kutta
method and the value obtained by homotopy-Pad\'{e} technique (see
\cite{li2}), where for briefly a few cases reported. Clearly, the
amplitude converges to the exact value for various $\epsilon$.

\begin{center}
\psfrag{w}[bc][t][1][180]{\large$a,\ \omega$}
\psfrag{h}[tl][t]{\large$h$}
\includegraphics{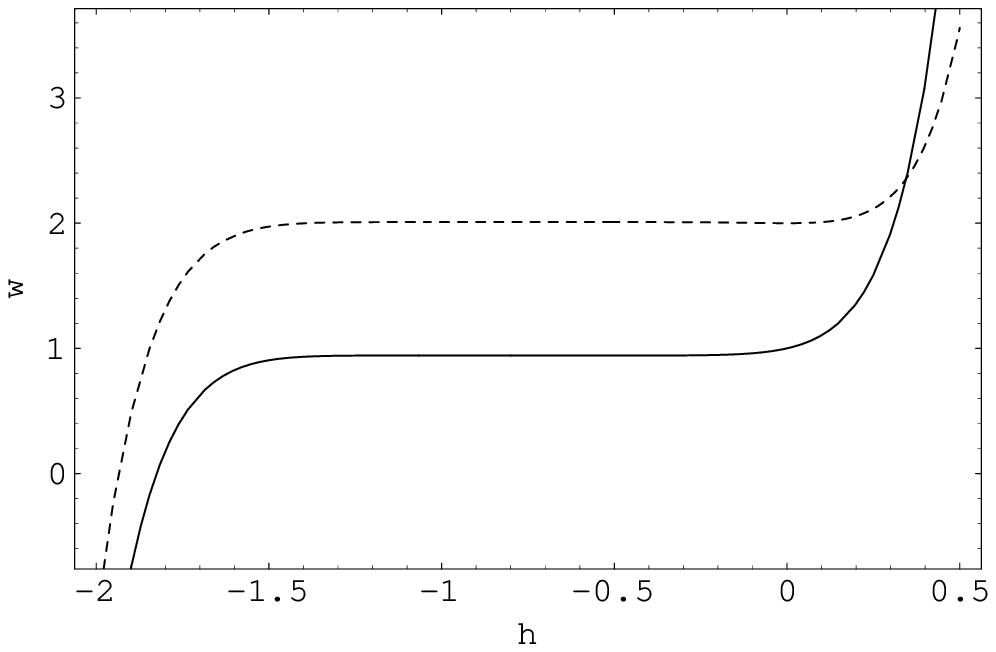}\\
\vspace{0.2cm} Fig.~1: The curves of the wave amplitude $a$ and
frequency $\omega$ versus $h$ for the 10th-order approximation
for $\epsilon=1$. Solid curve: the wave frequency; dotted line:
the wave amplitude.
\end{center}

\newpage
\begin{center}
\psfrag{w}[bc][t][1][180]{\large$a,\ \omega$}
\psfrag{h}[tl][t]{\large$h$}
\includegraphics{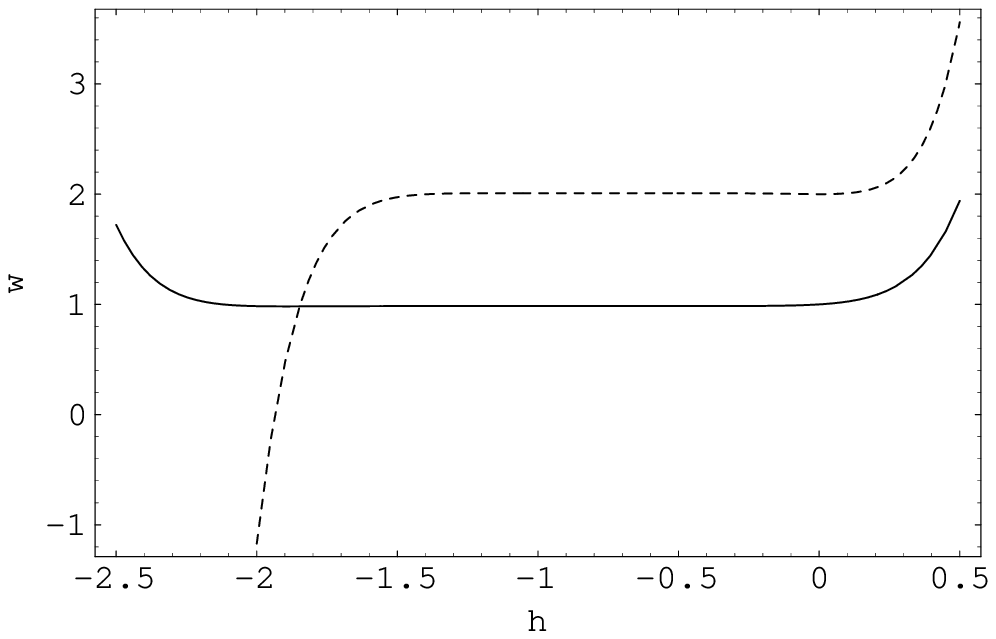}\\
\vspace{0.2cm} Fig.~2: The curves of the wave amplitude $a$ and
frequency $\omega$ versus $h$ for the 10th-order approximation
for $\epsilon=0.5$. Solid curve: the wave frequency; dotted line:
the wave amplitude.
\end{center}

\begin{center}
\psfrag{a}[bc][t][1][90]{\large$a$}
\psfrag{e}[tl][t]{\large$\epsilon$}
\includegraphics{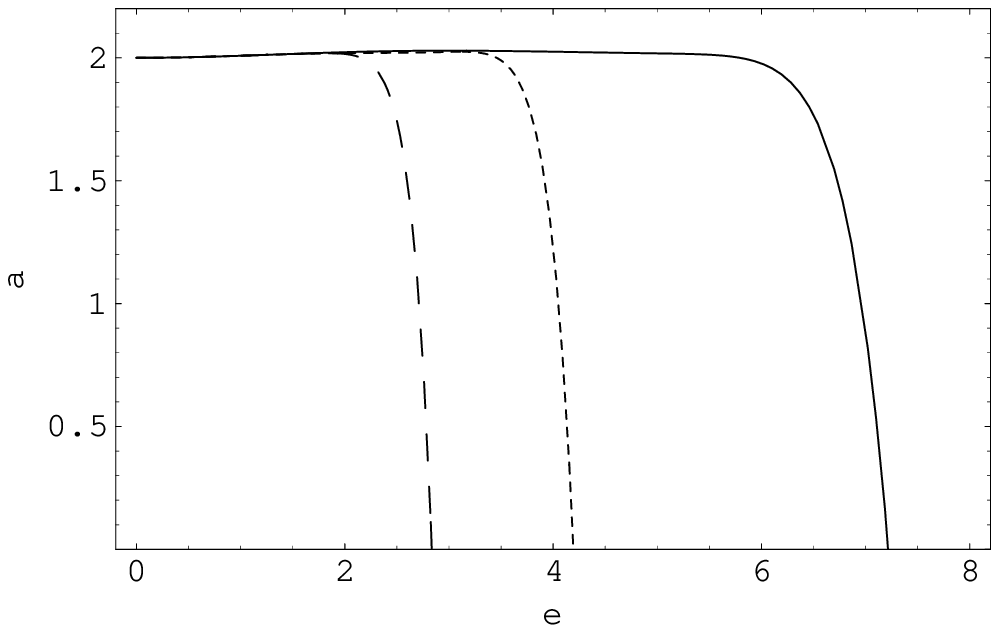}\\
\vspace{0.2cm} Fig.~3: Comparison of the amplitude of the
10th-order homotopy analysis approximation. Solid curve:
$h=-{1\over 3}$, dotted curve: $h=-{2\over 3}$, dashed
curve: $h=-1$.
\end{center}

\begin{center}
\psfrag{w}[bc][t][1][90]{\large$\omega$}
\psfrag{e}[tl][t]{\large$\epsilon$}
\includegraphics{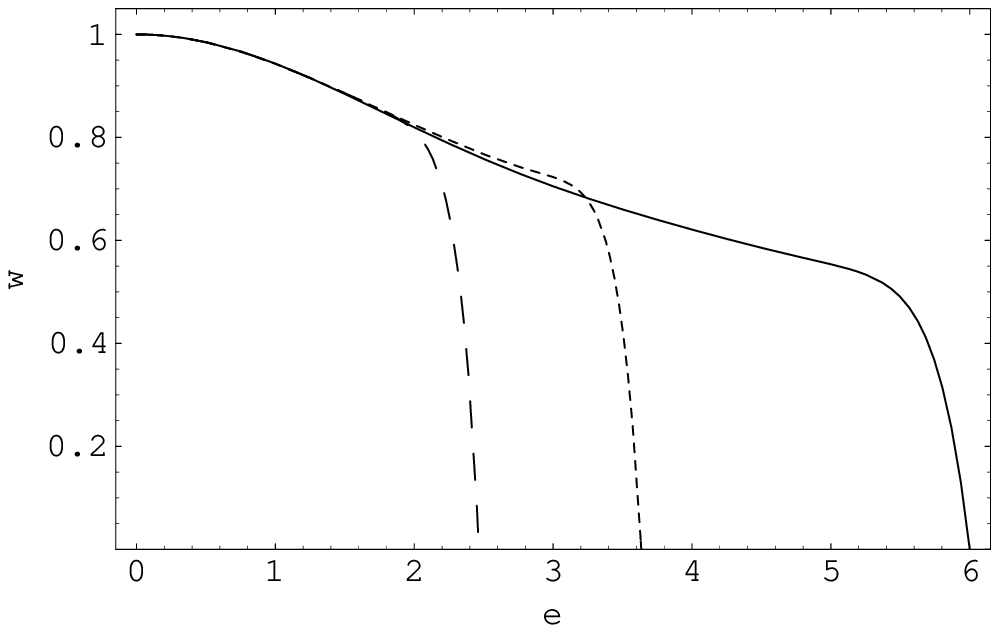}\\
\vspace{0.2cm} Fig.~4: Comparison of the frequency of the
10th-order homotopy analysis approximation. Solid curve:
$h=-{1\over 3}$, dotted curve: $h=-{2\over 3}$, dashed
curve: $h=-1$.
\end{center}

\begin{center}
\psfrag{a}[bc][t][1][90]{\large$a$}
\psfrag{e}[tl][t]{\large$\epsilon$}
\includegraphics{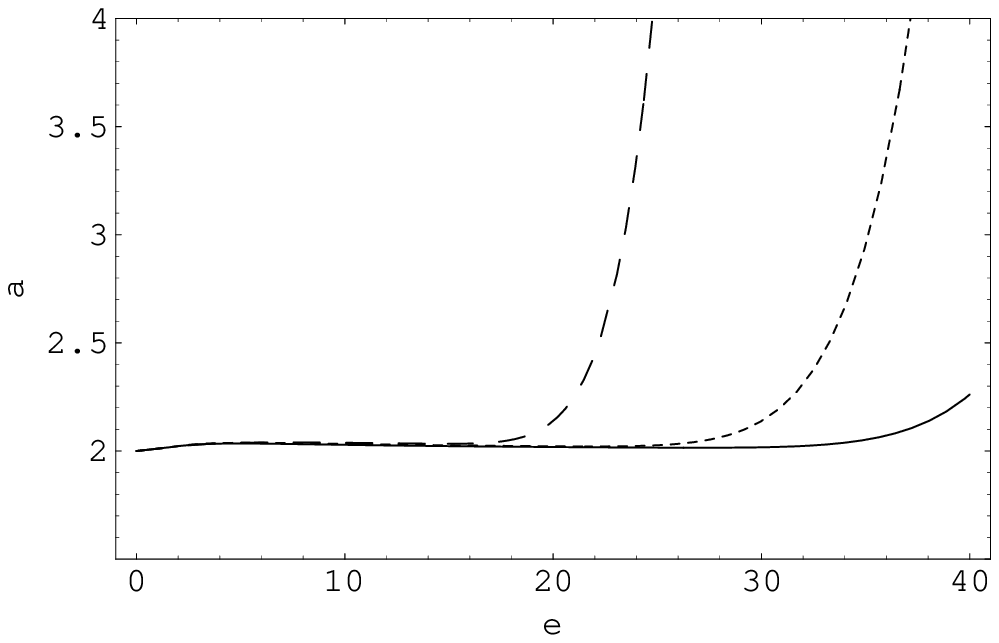}\\
\vspace{0.2cm} Fig.~5: Comparison of the amplitude of the
10th-order homotopy analysis approximation. Solid curve:
$\gamma=3$, dotted curve: $\gamma=2$, dashed curve: $\gamma=1$.
\end{center}

\newpage
\begin{center}
\psfrag{w}[bc][t][1][90]{\large$\omega$}
\psfrag{e}[tl][t]{\large$\epsilon$}
\includegraphics{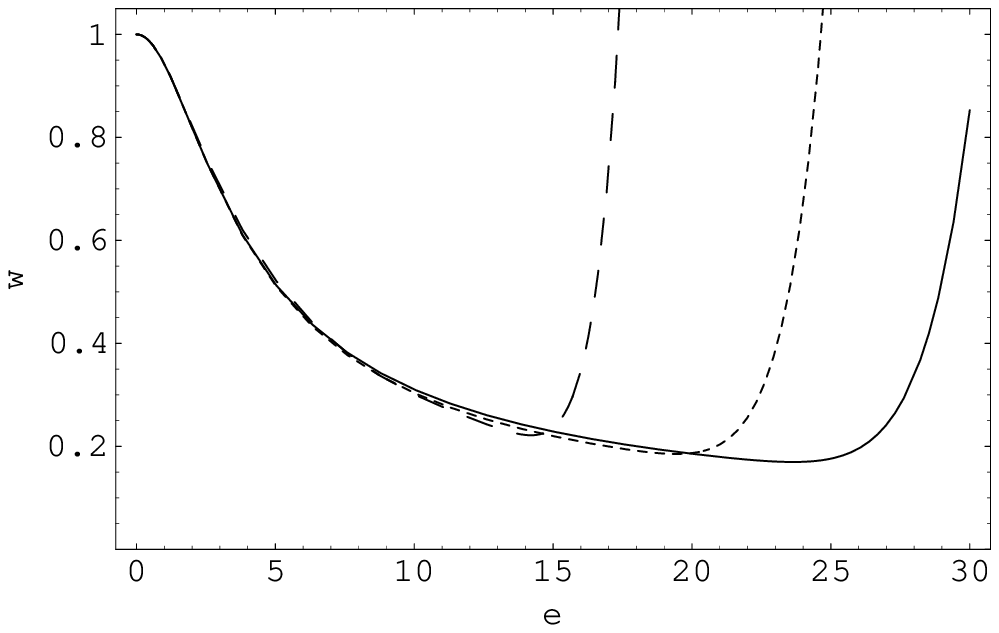}\\
\vspace{0.2cm} Fig.~6: Comparison of the frequency of the
10th-order homotopy analysis approximation. Solid curve:
$\gamma=3$, dotted curve: $\gamma=2$, dashed curve: $\gamma=1$.
\end{center}

\begin{center} Table 1: Results for $[m,m]$ Homotopy-Pad\'{e} approach for Example 1\\
\begin{tabular}{c|ccccccc}
\hline
 $\epsilon$ & 0.1 & 0.3 & 0.5 & 0.7 & 0.9 & 1.5 & 2.0 \\
 \hline
 $a_{RK}$ & 2.00010 & 2.00092 & 2.00248 & 2.00466 & 2.00724 & 2.01523 & 2.01989 \\
 $[2,2]$  & 2.00010 & 2.00092 & 2.00249 & 2.00469 & 2.00737 & 2.01670 & 2.02426 \\
 $[3,3]$  & 2.00010 & 2.00092 & 2.00249 & 2.00469 & 2.00737 & 2.01670 & 2.02427 \\
 $[4,4]$  & 2.00010 & 2.00092 & 2.00249 & 2.00466 & 2.00724 & 2.01515 & 2.01943 \\
 $[5,5]$  & 2.00010 & 2.00092 & 2.00249 & 2.00466 & 2.00724 &
 2.01514 & 2.01936 \\
 $[6,6]$  & 2.00010 & 2.00092 & 2.00249 & 2.00466 & 2.00724 &
2.01523 & 2.02001  \\
 $[7,7]$  & 2.00010 & 2.00092 & 2.00249 & 2.00466 & 2.00724 &
2.01523 & 2.02001  \\
 $[8,8]$  & 2.00010 & 2.00092 & 2.00249 & 2.00466 & 2.00724 &
2.01523 & 2.01989
\end{tabular} \end{center}

\noindent\textbf{Example 2.} Here, we consider $f(x)=5x^4-9x^2+1$.
This system has two limit cycles, one stable and the other one unstable
\cite{ll4}.

The corresponding approximation of their amplitudes,
\begin{eqnarray*}
a(\epsilon)& = & 1.755170 + 0.017880 \epsilon^2 + \cal{O}(\epsilon^4), \\
\bar{a}(\epsilon) & = & 0.720677+0.00390888\;\epsilon^2+{\cal O}(\epsilon^4),
\end{eqnarray*}
by a recursive algorithm was reported in \cite{ll2,ll3}. Under
transformation (\ref{trans}), Eq. (\ref{lie}) becomes
\begin{equation} \label{lie2}
\omega^2 u''(\tau) + \epsilon \omega \big[5a^4u^4(\tau) - 9a^2
u^2(\tau)+1 \big] u'(\tau) + u(\tau)=0.
\end{equation}
From (\ref{eq:11}), the term $R_m(\tau)$ in (\ref{mod}) becomes
\begin{eqnarray} \label{rm2}
R_m(\tau) & = & \sum_{n=0}^{m-1} u''_{m-1-n}(\tau) \Big(
\sum_{j=0}^n \omega_j \omega_{n-j} \Big) + u_{m-1}(\tau) +
\epsilon \sum_{n=0}^{m-1} \omega_n u'_{m-n-1}(\tau) \\
& & +5 \epsilon \sum_{n=0}^{m-1}\Big[ \Big( \sum_{i=0}^{m-1-n}
\omega_i u'_{m-n-i-1}(\tau) \Big) \sum_{j=0}^n \Big( \sum_{r=0}^j
\hat{a}_r \hat{a}_{j-r} \Big) \Big( \sum_{s=0}^{n-j}
\hat{u}_s(\tau) \hat{u}_{n-j-s}(\tau) \Big)\Big] \nonumber \\
\nonumber & & -9 \epsilon \sum_{n=0}^{m-1}\Big[ \Big(
\sum_{i=0}^{m-1-n} \omega_i u'_{m-n-i-1}(\tau) \Big) \sum_{j=0}^n
\Big( \sum_{r=0}^j a_r a_{j-r} \Big) \Big( \sum_{s=0}^{n-j}
u_s(\tau) u_{n-j-s}(\tau) \Big)\Big].  \nonumber
\end{eqnarray}
where
\begin{equation*}
\hat{a}_n=\sum_{i=0}^n a_i a_{n-i},\ \
\hat{u}_n(\tau)=\sum_{i=0}^n u_i(\tau) u_{n-i}(\tau).
\end{equation*}
It is found that the frequency $\omega$ and the amplitude $a$ at
the $M$th-order of approximation can be expressed in the form (\ref{res1}).
So, $a_0$ and $\omega_0$ are obtained by solving (\ref{ale}) for
$m=1$, i.e.
\begin{equation*}
c_{1,1}=(1-\omega_0^2)=0,\ \ \ d_{1,1}=\epsilon \omega_0 (8-18
a_0^2 + 5 a_0^4)=0.
\end{equation*}
Hence, we have two limit cycles with $\omega_0=1$: one of them with amplitude
${a}_0=\sqrt{{9+\sqrt{41} \over 5}}$ (stable limit cycle for
$\epsilon>0$), and the other one with amplitude $\bar{a}_0=\sqrt{{9-\sqrt{41} \over 5}}$
(unstable limit cycle for $\epsilon>0$). The obtained results for the
amplitude with $a_0$ as initial guess are as follows:
\begin{eqnarray*}
A_1 & = & 1.75517, \\
A_2 & = & 1.75517 + 0.0178803\,h^2\,{\epsilon }^2,\\
A_3 & = & 1.75517 + 0.0536409\,h^2\,{\epsilon }^2 +
0.0357606\,h^3\,{\epsilon }^2 + 0.0151888\,h^3\,{\epsilon }^4, \\
A_4 & = & 1.75517 + 0.107282\,h^2\,{\epsilon }^2 +
0.143042\,h^3\,{\epsilon }^2 + 0.0536409\,h^4\,{\epsilon }^2 +
  0.0607553\,h^3\,{\epsilon }^4 \\ & & - 0.179337\,h^4\,{\epsilon }^4 + 0.0129025\,h^4\,{\epsilon }^6.
\end{eqnarray*}
For the frequency are
\begin{eqnarray*}
\Omega_1 & = & 1 + 0.424737\,h\,{\epsilon }^2, \\
\Omega_2 & = & 1 + 0.849473\,h\,{\epsilon }^2 +
0.424737\,h^2\,{\epsilon }^2 + 0.270602\,h^2\,{\epsilon }^4,
\\
\Omega_3 & = & 1 + 1.27421\,h\,{\epsilon }^2 +
1.27421\,h^2\,{\epsilon }^2 + 0.424737\,h^3\,{\epsilon }^2 +
0.811805\,h^2\,{\epsilon }^4 \\ & & +
  0.451679\,h^3\,{\epsilon }^4 + 0.191558\,h^3\,{\epsilon }^6
, \\
\Omega_4 & = & 1 + 1.69895\,h\,{\epsilon }^2 +
2.54842\,h^2\,{\epsilon }^2 + 1.69895\,h^3\,{\epsilon }^2 +
0.424737\,h^4\,{\epsilon }^2 +
  1.62361\,h^2\,{\epsilon }^4 \\ & & + 1.80672\,h^3\,{\epsilon }^4 + 0.543231\,h^4\,
  {\epsilon }^4 + 0.76623\,h^3\,{\epsilon }^6 +
  0.38455\,h^4\,{\epsilon }^6 + 0.142383\,h^4\,{\epsilon }^8.
\end{eqnarray*}
In particular, for $h=-1$, in 10th-order approximation we obtain
\begin{eqnarray*}
A_{10} & = & 1.75517 + 0.0178803\,{\epsilon }^2 -
0.240092\,{\epsilon }^4 + 0.859582\,{\epsilon }^6 -
0.227156\,{\epsilon }^8 -
  13.7118\,{\epsilon }^{10} \\
  & & + 10.9555\,{\epsilon }^{12} + 4.73704\,{\epsilon }^{14} - 0.626997\,{\epsilon }^{16} +
  0.00484811\,{\epsilon }^{18} + \cal{O}(\epsilon^{20}).
\end{eqnarray*}
And for $h=-1$, in 10th-order approximation with $\bar{a}_0$
as initial guess, we have
\begin{eqnarray*}
\bar{A}_{10} & = & 0.720677 + 0.00390888\,{\epsilon }^2 -
0.000410295\,{\epsilon }^4 - 0.0000165055\,{\epsilon }^6 \\ & & +
  9.11444\,{10}^{-6}\,{\epsilon }^8
  - 1.45045\,{10}^{-7}\,{\epsilon }^{10} - 2.42445\,{10}^{-7}\,{\epsilon }^{12} -
  1.45593\,{10}^{-7}\,{\epsilon }^{14}\\ & &  + 9.91724\,{10}^{-10}\,{\epsilon }^{16}
  + 5.37168\,{10}^{-11}\,{\epsilon }^{18} +
\cal{O}(\epsilon^{20}).
\end{eqnarray*}

We can investigate the influence of $h$ on the convergence of
$a$ and $\omega$ by plotting the curve of $a$ and $\omega$ versus
$h$, as shown in Fig.~7. One can see that, for $\epsilon=0.5$,
we have $-0.9\leq h \leq -0.2$. The comparison of the
amplitude $a$ and the frequency $\omega$ at the 10th-order of
approximation with the numerical results is as shown in Figs.~8
and 9, where $h=-1,\ -{2\over 3}$ and $-{1\over 3}$. However,
as $h$ is negative and close to zero, the convergence region
becomes larger and larger, as in Example 1.

\begin{center}
\psfrag{w}[bc][t][1][180]{\large$a,\ \omega$}
\psfrag{h}[tl][t]{\large$h$}
\includegraphics{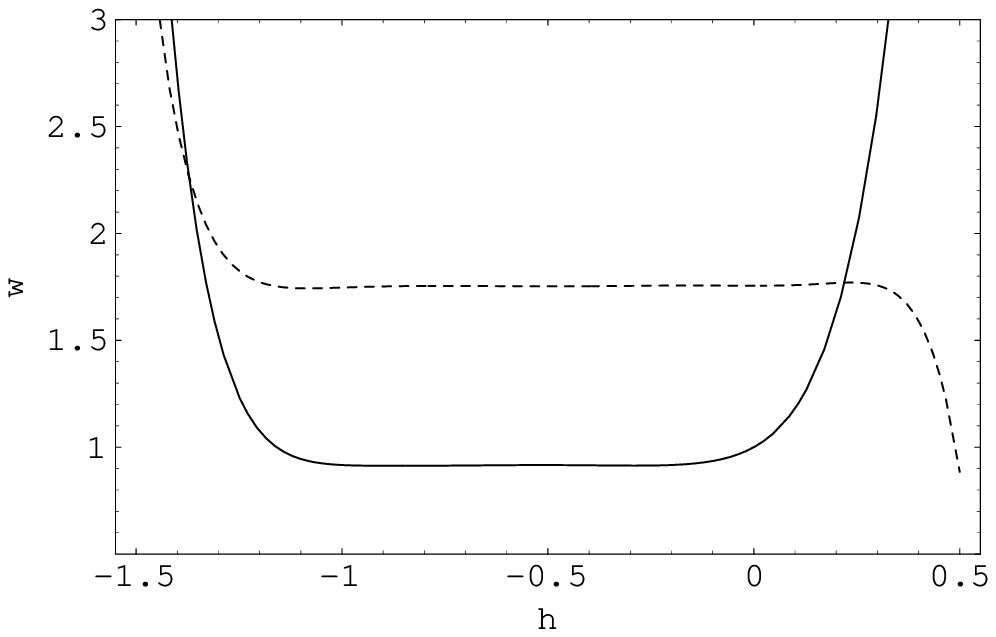}\\
\vspace{0.2cm} Fig.~7: The curves of the wave amplitude $a$ and
frequency $\omega$ versus $h$ for the 10th-order approximation
for $\epsilon=0.5$. Solid curve: the wave frequency; dotted line:
the wave amplitude.
\end{center}

\begin{center}
\psfrag{a}[bc][t][1][90]{\large$a$}
\psfrag{e}[tl][t]{\large$\epsilon$}
\includegraphics{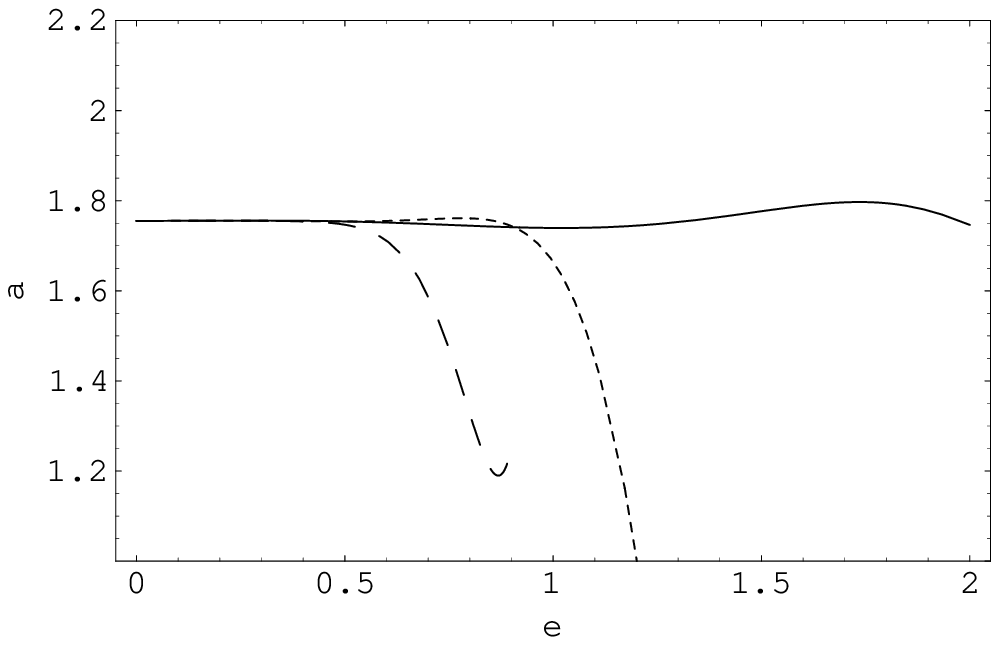}\\
\vspace{0.2cm} Fig.~8: Comparison of the amplitude of the
10th-order homotopy analysis approximation. Solid curve:
$h=-{1\over 3}$, dotted curve: $h=-{2\over 3}$, dashed
curve: $h=-1$.
\end{center}

\begin{center}
\psfrag{w}[bc][t][1][90]{\large$\omega$}
\psfrag{e}[tl][t]{\large$\epsilon$}
\includegraphics{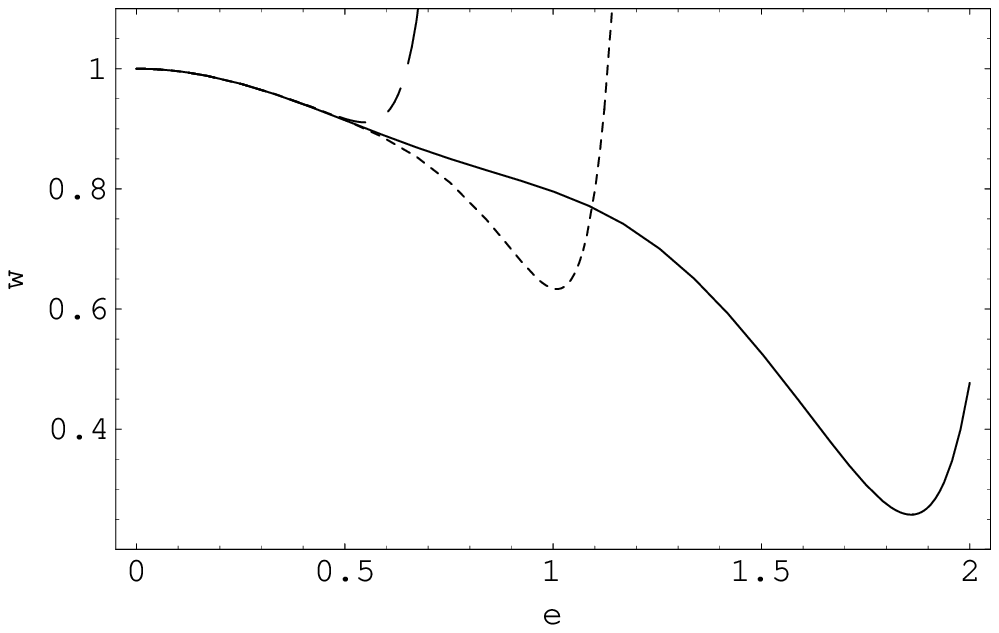}\\
\vspace{0.2cm} Fig.~9: Comparison of the frequency of the
10th-order homotopy analysis approximation. Solid curve:
$h=-{1\over 3}$, dotted curve: $h=-{2\over 3}$,
dashed curve: $h=-1$.
\end{center}

\section{Conclusions}
\setcounter{equation}{0}

We have applied the homotopy analysis method (HAM) to the
classical Li\'{e}nard differential equation (\ref{lie}) to obtain
analytic approximations of the amplitude and frequency of its limit cycles.
Two examples have been explicitly worked out. The results obtained
with the HAM are in excellent agreement with the known solutions.
Moreover, the HAM provides us with a convenient way (the parameter $h$) to control the
convergence of approximation series; this is a fundamental
qualitative difference between the HAM and other methods for
finding approximate solutions. In particular, the case $h=-1$
corresponds with the exact perturbative expansion in $\epsilon$.

Let us conclude by saying that the examples shown in this paper are
illustrative of the power of the HAM to solve complicated
nonlinear problems.

\end{document}